\documentclass[12pt]{article}

\usepackage{epsfig}
\usepackage{amsmath}
\usepackage{verbatim}

\newcommand{\nc}{\newcommand}
\nc{\be}{\begin{equation}}
\nc{\ee}{\end{equation}}
\nc{\bea}{\begin{eqnarray}}
\nc{\eea}{\end{eqnarray}}
\nc{\nn}{\nonumber}

\nc{\lp}{\left(}
\nc{\rp}{\right)}

\nc{\rme}{{\textrm{e}}}

\nc{\markb}{$\clubsuit \Rightarrow $}
\nc{\marke}{$ \Leftarrow \clubsuit $}
\nc{\markx}{$ \clubsuit !! \clubsuit $}
\nc{\markeq}{ \clubsuit !! \clubsuit }

\nc{\eq}{{Eq.}}
\nc{\eqs}{{Eqs.}}

\begin{document}

\begin{flushright}
{\tt MAN/HEP/2006/29}
\end{flushright}

\begin{center}

{\Large\bf Hybrid Inflation Exit through Tunneling}

\vskip 1cm

Bj\"orn Garbrecht\\
{\it
School of Physics \& Astronomy,
The University of~Manchester,\\
Oxford~Road,
Manchester M13~9PL, United Kingdom}

\vskip .5cm

Thomas Konstandin
\\{\it
Department of Theoretical Physics,
Royal Institute of Technology (KTH),\\
AlbaNova University Center,
Roslagstullsbacken 21,\\
106 91 Stockholm, Sweden}

\end{center}

\vskip .5cm

\begin{abstract}
For hybrid inflationary potentials, we derive the tunneling rate from
field configurations along the flat direction towards the waterfall regime.
This process competes with the classically
rolling evolution of the scalar fields and needs to be strongly
subdominant for
phenomenologically viable models. Tunneling may exclude models with a mass
scale below $10^{12}\, {\rm GeV}$, but can be suppressed by small values of the
coupling constants. We find that tunneling is negligible for
those models, which do not require fine tuning in order to cancel radiative
corrections, in particular for GUT-scale SUSY inflation.
In contrast, electroweak scale hybrid inflation is not viable,
unless the inflaton-waterfall field coupling is smaller than
approximately $10^{-11}$.
\end{abstract}

\section{Introduction}

The slow roll paradigm of inflation~\cite{LythRiotto} requires the
scalar potential to be flat to such an extent, that the Hubble
expansion causes an overdamping of the evolution of the inflaton
field. This has the consequence, that the kinetic energy of the
inflaton is negligible, and the equation of state of the dominant
component of the Universe is approximately the same as for vacuum
energy.  When realizing slow roll inflation within single field
models, one encounters the problem of reconciling the flatness of the
potential, its comparably large magnitude and the wish to keep the
vacuum expectation value (VEV) of the inflaton below the Planck scale.

A possibility to address this problem is the hybrid inflation~\cite{Linde}
mechanism, where the slowly rolling inflaton
triggers a waterfall field to rapidly roll down the
potential and to terminate inflation  at some critical point.
The direction along which the
potential increases towards large values driving inflation and the
direction of the slow-roll are therefore separated.

When comparing different models of hybrid inflation at the same scale, that is
with the same value
of the potential, it is clear that in a model which has a flatter
direction for the inflaton, a certain comoving scale leaves the horizon
when the inflaton is closer to the critical point than in a model with a
steeper direction. Imagining the limiting case of a completely flat direction,
the classical field dynamics suggest that inflation may last infinitely
long with the inflaton being arbitrarily close to the critical point.
However, within quantum theory, metastable configurations eventually always
decay to the one of lowest energy. We therefore expect that in hybrid
inflation, a field configuration along the flat direction may tunnel
to form a bubble containing a field configuration in which inflation
ends and the scalar fields rapidly assume the true vacuum state.
It is the purpose of this study, to estimate this decay rate, compare
it to the classical field evolution and to specify for which model
parameters tunneling is a non-negligible effect.

\section{Tunneling during Inflation}

\subsection{Tunneling without Barriers}

The semiclassical theory of tunneling for scalar field theory is developed
by Coleman and Callan in~\cite{Coleman,CallanColeman}.
We consider the Lagrangian
\begin{equation}
{\cal L}=\frac 12 (\partial_\mu \varphi)(\partial^\mu\varphi) -V(\varphi)\,,
\end{equation}
where the field $\varphi$ is initially located everywhere in space at the point
$\varphi_+$, which corresponds to
a false vacuum  or a classically metastable configuration, and where
we normalize $V(\varphi_+)=0$.
In order to calculate the decay rate, one proceeds by solving the
classical Euclidean equation of motion
\begin{equation}
\label{EOM:Eu}
\frac{\partial^2\varphi}{\partial \varrho^2}+\frac{3}{\varrho}
\frac{\partial\varphi}{\partial \varrho} = V^\prime(\varphi)\,,
\end{equation}
where the prime denotes a derivative with respect to $\varphi$. We assume
that the solution takes a spherical symmetric form in Euclidean space and
write $\varrho=|x|$. In order to understand the properties of the solutions
to this equation, it is most useful to recall that it corresponds to
the equation of motion for a one-dimensional particle
moving in the potential $V(\phi)$ turned upside down and with a friction term
$(3/\varrho)(\partial\varphi/\partial\varrho)$, which implies infinite
damping at $\varrho=0$ and vanishing damping when $\varrho \to \infty$.

The instanton solution, which obeys the boundary condition
$\varphi(\infty)=\varphi_+$ is called the bounce, and we denote it by
$\overline \varphi(\varrho)$. It uniquely determines a release point
$\varphi_{\rm r}$, at which $V(\varphi_{\rm r})<V(\varphi_+)$ and which
satisfies $\partial \overline \varphi / \partial \varrho =0$ at
$\overline \varphi=\varphi_{\rm r}$ and $\varrho=0$.
Physically, $\varphi_{\rm r}$ is the
initial value of the scalar field inside a nucleating bubble, from which
it starts to evolve classically.

Having found the bounce solution, we can compute its Euclidean action
\begin{equation}
\label{S:Eu}
S_{\rm E}=2 \pi^2 \int\limits_0^\infty \varrho^3 d\varrho
\left[\frac 12 \left(\frac{\partial\overline\varphi}{\partial\varrho}\right)^2+V(\overline\varphi)\right]\,,
\end{equation}
which is used to obtain the tunneling rate $\Gamma$ per volume ${\cal V}$ as
\begin{equation}
\label{tunnelingrate}
\frac \Gamma {\cal V} = \frac{S_{\rm E}^2}{4\pi^2}
\left(
\frac{\det^\prime\left[-\partial^2+V^{\prime\prime}(\overline \varphi)\right]}{\det\left[-\partial^2 + V^{\prime\prime}(\varphi_+)\right]}
\right)^{-1/2}{\rm e}^{-S_{\rm E}}\,,
\end{equation}
where the prime at the determinant indicates the omission of the zero
eigenvalues. The evaluation of the determinants is a quite costly task,
and we follow the common
practice~\cite{WeinbergLee,LindeTunnel,FelderKofmanLinde} to estimate
their
values from the parameters of the particular theory under consideration.
Indeed, the results we present justify this procedure a posteriori.

We intend to apply this theory of tunneling to
hybrid inflation, which is implemented by the generic potential~\cite{Linde}
\begin{equation}
\label{hybpot}
V(\sigma,\phi)= V_0(\sigma,\phi) +  V_L(\sigma)
= \frac\lambda 4 \phi^4 -  \frac{m^2}2  \phi^2
+ \frac{m^4}{ 4\lambda}  + \frac12 g^2 \phi^2 \sigma^2 + V_L(\sigma)\,.
\end{equation}
This potential is almost flat with respect to the inflaton
$\sigma$ along the direction
where $\phi=0$. The flat direction is lifted by the
contribution $V_{\rm L}(\sigma)$,
where we normalize $V_{\rm L}(0)=0$,
which causes $\sigma$ to classically roll down the potential from
larger to smaller values. Inflation ends shortly after
$\sigma$ reaches the critical value
\begin{equation}
\sigma_{\rm c}=\frac mg \,.
\end{equation}
At this point, the mass square for the field $\phi$ changes its sign from
positive to negative and the inflationary valley turns into a ridge.
The field $\phi$ then quickly evolves away from zero
and the fields eventually assume the values
\begin{eqnarray}
\label{glmin}
\sigma_0=0\,,\quad \phi_0= \frac{m}{\sqrt{\lambda}} \,,
\end{eqnarray}
where $V(\sigma_0,\phi_0)=0$ and inflation is terminated.
Due to the transition from valley to ridge, from which the fields
fall, this is called the waterfall
mechanism, and we denote the area where $\sigma<\sigma_{\rm c}$ as the
waterfall region.

Returning to the question of tunneling, we note that
the hybrid potential~(\ref{hybpot}) does not have any local minima
but the global one~(\ref{glmin}). Therefore, there are no false vacuum
configurations possible, and it may appear that the
theory of tunneling and bubble nucleation does not play any role for
hybrid scenarios. However, as already mentioned in the introduction,
we can imagine the case $V_{\rm L}(\sigma)=0$ and wonder whether
a configuration with $\sigma>\sigma_{\rm c}$ is stable.

Quite similar situations are discussed by Weinberg and Lee~\cite{WeinbergLee},
and they point out that a bounce solution can exist in some
cases without a potential barrier between the initial point and the
global minimum of the potential\footnote{
Linde has given an earlier example of upside-down $\phi^4$-theory,
where tunneling can occur~\cite{LindeTunnel}. See
also~\cite{FelderKofmanLinde}
for a more recent related discussion.
}.
The necessary condition for the existence of a bounce is not the
presence of a potential barrier, but of an energy barrier,
constituted by the potential and a contribution from the gradient terms
of the bubble wall.
Therefore, a false vacuum is not required
to exist for tunneling to be a relevant process.
A very instructive example is given by the potential
\be
V(\phi) = \left\{ 
\begin{aligned}
0, \quad &\phi < 0 \\
-k\phi \quad & \phi \geq 0 \\
\end{aligned}
\right.\, .
\ee
Classically, if the field is positioned on the plateau at the position
$\phi=-\Delta \phi$, the system
would be stable, while quantum-mechanically, it turns out to be unstable
due to tunneling. 
The existence of a corresponding bounce solution can be understood from
the Euclidean equation of motion~(\ref{EOM:Eu}).
If the field is released at rest when $\varrho=0$
and $\phi=\phi_{\rm r}>0$, it
will accelerate in the upside-down potential until $\phi=0$ and then
asymptotically come
to rest again at $\phi=-\Delta \phi$ due to the damping term. This
bounce solution therefore describes tunneling from the metastable position
$\phi=-\Delta \phi$ on the plateau to nucleate a bubble with the
vacuum expectation value $\phi=\phi_{\rm r}$ inside. The rate for
this to happen is calculated to be~\cite{WeinbergLee}
\begin{equation}
\label{rateWL}
\frac{\Gamma}{\cal V}=
C\frac 49 \pi^2 \Delta \phi^4
\exp\left(-\frac{32\pi^2}3 \frac{\Delta\phi^3}k \right)
\,,
\end{equation}
where $C$ is a constant of order one, which can in principle be determined
by evaluating the determinants in \eq~(\ref{tunnelingrate}).
This result is apparently already very useful in order to estimate
whether for a given inflationary model,
it is in order to worry about tunneling. If the cube of the distance from
the region where inflation takes place to some other point of lower potential
is of the same order or smaller than the derivative of the potential at
that point, the bounce action can be of order one and tunneling sizeable.
Similar to this example, for the hybrid potential $V_0$,
\eq~(\ref{hybpot}), bounce solutions exist that start in the waterfall
region and come at rest on the flat direction where $\phi=0$ and
$\sigma>\sigma_c$.

One may argue that the potential during inflation is not exactly flat and
that therefore the formula~(\ref{rateWL}) for the tunneling rate does
not apply. We follow however the argument of Weinberg and Lee,
that taking the motion of the inflaton field or the lifting of the flat
direction into account will only reduce the action of the tunneling process.
For calculating the bubble nucleation rate in the hybrid model,
we therefore determine the bounce
solution for the potential $V_0$ and neglect the effect of $V_{\rm L}$.
This way, we obtain a lower bound for the tunneling probability, which still
allows to derive constraints on the parameter space for hybrid inflation.

\subsection{Numerical Results}

We now determine the bounce action for the hybrid potential~(\ref{hybpot})
as a function of the distance of the inflaton from the critical point,
\begin{equation}
\Delta\sigma=\sigma-\sigma_{\rm c}\,.
\end{equation}
While it is not possible to find analytic bounce solutions, one can
reduce the problem considerably by making use of the scaling properties of
the potential.
Inspecting the Euclidean equations of motion~(\ref{EOM:Eu})
for the hybrid case,
\begin{eqnarray}
\frac{\partial^2 \sigma}{\partial \varrho^2}
+\frac 3\varrho \frac{\partial \sigma}{\partial \varrho}\!\!\!&=&\!\!\!
g^2 \phi^2 \sigma\,, \nonumber\\
\frac{\partial^2 \phi}{\partial \varrho^2}
+\frac 3\varrho \frac{\partial \phi}{\partial \varrho}\!\!\!&=&\!\!\!
-m^2 \phi + g^2 \sigma^2 \phi + 4\lambda \phi^3
\label{EoMexplicit}
\,,
\end{eqnarray}
we see that they are left invariant under the following rescaling:
\be
\lambda \to \lambda \kappa, \quad \rho \to \rho \kappa^{-1/2}, \quad
m \to m \kappa^{1/2},\quad g \to g \kappa^{1/2}\,.
\ee
The bounce action~(\ref{S:Eu}) then transforms as
\be
S_E(\lambda,m,g, \Delta \sigma / \sigma_c) = \lambda^{-1}
S_E(1,m/\sqrt{\lambda},g/\sqrt{\lambda}, \Delta \sigma / \sigma_c)\,.
\ee
Another rescaling leaving the equations of motion~(\ref{EoMexplicit})
invariant is
\be
m \to \kappa m, \quad \sigma \to \kappa \sigma, 
\quad \phi \to \kappa \phi, \quad \rho \to \kappa^{-1} \rho\,.
\ee
This reveals that $S_E(\lambda,m,g, \Delta \sigma / \sigma_c)$
does not depend on $m$,
\be
S_E(\lambda,m,g, \Delta \sigma / \sigma_c)
= \lambda^{-1} S_E(1,m_0,g/\sqrt{\lambda}, \Delta \sigma / \sigma_c) 
=: \lambda^{-1} \chi (g/\sqrt{\lambda}, \Delta \sigma / \sigma_c)\,,
\label{SE:para}
\ee
where $m_0$ is arbitrary.

We now determine the function $\chi$ numerically. 
In general, finding bounce solutions can be very complicated for multi-dimensional problems, or at least time consuming. Two
algorithms, that can be applied to a wide range of problems, have been
presented, {\it e.g.} in Refs.~\cite{Konstandin:2006nd, Cline:1999wi}.
These algorithms are not immediately applicable to our problem, since
they have been designed for the case of tunneling with potential
barriers. Fortunately, for two-dimensional problems, one can resort to
scan procedures, which we apply here. First, we fix the
starting point of the configuration $(\sigma_0, \phi_0)$ and solve the
equations of motion by integration. For late times, the solution can behave
in two qualitatively different ways.
The first possibility is that $\sigma$ always stays
smaller than $\sigma_c$, and $\phi$ oscillates around zero. In this case
$\sigma_0$ was chosen too small.  In the second case, $\sigma$
is finally larger than $\sigma_c$ and the upside-down
potential is hence unstable
in the $\phi$-direction. Depending on the initial point, the
configuration then behaves usually as $\phi \to \pm \infty$, when
$\rho \to
\infty$. These two cases correspond to the 'over-/undershooting'
of the one-dimensional problem. Keeping $\phi_0$ fixed, while varying
$\sigma_0$ using the 'over-/undershooting' method, leads thus to a
bounce solution.

\begin{figure}[t]
\begin{center}
\epsfig{file=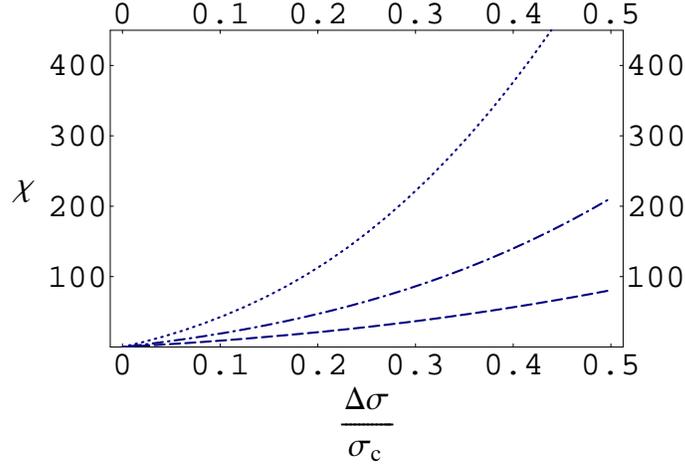, width=4.0in} 
\end{center}
\caption{
\label{SE_2d}
\small
The two-dimensional numerical result for the function
$\chi(g/\sqrt{\lambda}, \Delta \sigma / \sigma_c)$ for the values
$g/\sqrt{\lambda}=\sqrt{2}$ (dashed),
$g/\sqrt{\lambda}=1$ (dot-dashed) and 
$g/\sqrt{\lambda}=1/\sqrt{2}$ (dotted).
}
\end{figure}

In Fig.~\ref{SE_2d}, we plot the function
$\chi(g/\sqrt{\lambda}, \Delta \sigma / \sigma_c)$, obtained
by the above procedure, for different values of 
$g/\sqrt{\lambda}$.
The results show for small $\Delta \sigma\ll m g/\lambda$ a scaling according
to $g^{-2}  \Delta\sigma$, which we explain below.
The numerical coefficient turns out to be
\be
\label{SE:2d}
S_{\rm E} \approx 158 \times \frac{\Delta \sigma}{\sigma_c} \frac{1}{g^2}.  
\ee

To ensure this scaling behaviour even for very small values for the
coupling constants, we will give an analytical upper bound for the
Euclidean action in the following. With above insights, we can proceed
with further simplifications of the problem.  Sizeable tunneling may
only occur when the inflaton $\sigma$ is close to its critical value,
{\it cf.} Fig.~\ref{SE_2d}.
Therefore, we assume 
\be
\label{assume1}
\Delta\sigma \ll \sigma_{\rm c}.
\ee
In order to obtain a lower bound on the tunneling rate, we impose the
instanton to follow a straight trajectory in $(\sigma,\phi)$ space.
The exact solution along a curved trajectory has a lower Euclidean
action and therefore corresponds to a larger tunneling rate.
The trajectory is parameterized by
\begin{eqnarray}
\phi&=&aw\nonumber\,,\\
\sigma&=&\Delta\sigma+\frac mg - \sqrt{1-a^2}w\,,
\end{eqnarray}
where $a\in[0;1]$ is a free parameter that will be determined by
minimizing the action. Along this trajectory, the
potential~(\ref{hybpot}) close to the critical point takes the form
\begin{equation}
\label{V:approx}
V=\frac 14 \frac{m^4}{\lambda}-a^2 \sqrt{1-a^2}gmw^3+a^2gm\Delta\sigma w^2
+O\left(w^2\Delta \sigma^2,\;w^4\right)
\,.
\end{equation}
We now determine the value of the parameter $a$, for which the Euclidean bounce
action is minimal. For that purpose, we consider the potential
\begin{equation}
V=-\alpha w^3 +\beta w^2\,.
\end{equation}
By rescaling arguments, one obtains that 
the corresponding action has to scale as
\be
S_{\rm E} \sim \frac{\beta}{\alpha^2} \sim \frac{\Delta \sigma}{ m \, g}
 \times \frac1{a^2(1-a^2)}
\ee
and is minimized for $a=1/\sqrt{2}$. This explains the scaling behaviour for
small $\Delta \sigma$ observed in~(\ref{SE:2d}).
The comparison with
\eq~(\ref{SE:para}) yields for the linearized case
\bea
S_E &\sim& \quad \frac{\Delta \sigma}{\sigma_c} \frac{1}{g^2}\,,  \\
\chi (g/\sqrt{\lambda}, \Delta \sigma / \sigma_c)
&\sim& \lp\frac{g}{\sqrt{\lambda}}\rp^{-2} \times \frac{\Delta \sigma}{\sigma_c}\,.
\eea
For the choice $a=1/\sqrt 2$,
neglecting the $w^4$ terms in the approximated linearized
potential~(\ref{V:approx}) is justified when
\be
\label{assume2}
w\ll \sqrt{32} \sigma_{\rm c} \frac{g^2}{\lambda}\,,
\ee
and the $w^2\Delta\sigma^2$ terms are subdominant if
\be
\label{assume3}
\Delta\sigma \ll 2 \sigma_{\rm c}\,.
\ee
Numerically, we find for the constant of
proportionality
\be
\label{SENum}
S_{\rm E}= 182\times \frac{\Delta \sigma}{\sigma_c} \frac{1}{g^2}\,,
\ee
where the larger factor of proportionality when compared with~(\ref{SE:2d}) is
due to the fact that we are restricted to the linear path and therefore
miss the minimum of the Euclidean action in the two-dimensional field space.
We also note that the point $w_{\rm r}$, from which the field $w$
in the bounce solution is released, scales according to
\begin{equation}
\label{wr}
w_{\rm r}=
8.2 \times \frac{\Delta\sigma}{\sigma_{\rm c}}\frac mg
=8.2 \times \Delta \sigma\,.
\end{equation}
Notice that a small Euclidean action, $S_{\rm E} \ll 1$, automatically
ensures the requirements in \eqs~(\ref{assume2}) and (\ref{assume3})
and hence the validity of the approximation in \eq~(\ref{V:approx}),
if $g,\lambda < 1$. 

Finally, when assuming
$m$ to be of order of the Grand Unified Scale $10^{16}\,{\rm GeV}$ or less,
all scales in the problem are larger than the Hubble rate\footnote{
The displacement $\Delta \sigma$ exceeds the Hubble rate as a consequence
of imposing the small observed value~(\ref{SPRnum}) on the
the amplitude of the scalar perturbations~(\ref{SPR}).
}
\begin{equation}
\label{Hubble}
H=\sqrt{\frac{8\pi V}{3 m_{\rm Pl}^2}}\,,
\end{equation}
where $m_{\rm Pl}=1.22 \times 10^{19} \,{\rm GeV}$ denotes the Planck mass,
such that
gravitational effects can be neglected~\cite{ColemanDeLuccia}.

\section{Bounds on Specific Models}

We estimate the relevant values for $\Delta \sigma$ using
the standard slow-roll dynamics of the inflaton.
When the expectation value of the inflaton,
at a certain instant during inflation, takes the value
$\sigma=\sigma_{\rm e}$, the number of e-foldings $N_{\rm e}$ that will
elapse until inflation ends is calculated as
\begin{equation}
\label{Ne}
N_{\rm e}=\int\limits_{\sigma_{\rm e}}^{\sigma_{\rm c}} H\,dt
=\frac{8\pi}{m_{\rm Pl}^2}V \int\limits_{\sigma_{\rm c}}^{\sigma_{\rm e}}
\frac{d\sigma}{\partial V/\partial \sigma}\,,
\end{equation}
where we have used the slow-roll approximation
$3H \partial \sigma / \partial t =-\partial V /\partial \sigma$.
One important observational constraint is the amplitude $\sqrt{P_{\cal R}}$
of the power spectrum of scalar perturbations
for the scale $k$, that exits the horizon when
$\sigma=\sigma_{\rm e}$,
\begin{equation}
\label{SPR}
\sqrt{P_{\cal R}}=\sqrt\frac \pi 6 \frac {16}{m_{\rm Pl}^3}
\frac{V^{3/2}}{\partial V / \partial \sigma}\,\Bigg|_{\sigma=\sigma_{\rm e}}\,.
\end{equation}
Here, we impose the normalization~\cite{WMAP3}
\begin{equation}
\label{SPRnum}
\sqrt{P_{\cal R}}=4.5 \times 10^{-5}
\end{equation}
at $k=0.05 \, {\rm Mpc^{-1}}$.
This scale exits the horizon at
\begin{equation}
N_{\rm e}=50+\frac 13 \log_{10} \frac{T_{\rm R}}{10^9 \, {\rm GeV}}
+\frac 23 \log_{10} \frac{V^{1/4}}{10^{15} \, {\rm GeV}}
\,.
\end{equation}
Since $k=0.05 \, {\rm Mpc^{-1}}$ corresponds to multipole moments around
$\ell=700$, the largest angular observable scales have exited the horizon
about six to seven e-folds earlier.

A very conservative estimate for $\Delta \sigma$ and therefore the tunneling
rate is therefore obtained by setting $N_{\rm e}=60$ and
\begin{equation}
\Delta \sigma = \sigma_{\rm e}-\sigma_{\rm c}\,.
\end{equation}
We use this value to compute the Euclidean action~(\ref{SENum}) and to
estimate the tunneling rate~(\ref{tunnelingrate}). The latter is to be compared
with the expansion rate during inflation $H$, {\it e.g.} the number
of non-inflationary bubbles nucleated per expansion time in one horizon
is given by $\Gamma/({\cal V}H^4)$ and should be much less than one. An
interesting, but difficult question would be to quantify how much less.
Due to the exponentially strong dependence of the tunneling rate on
the model parameters, we omit a discussion of this question by the
same token on which we do not evaluate the determinants in
\eq~(\ref{tunnelingrate}).

We furthermore remark that it appears very likely that for viable
inflationary models,
one has to impose that tunneling also does not occur at much lower values of
$N_{\rm e}$ than 60.
The nucleation of non-inflationary bubbles would lead to very
large density perturbations on small scales, which induce the production
of primordial black holes~\cite{Carr}, which is strongly constrained
observationally~\cite{GreenLiddle}. We do not discuss this possibility here
any further and just explore the conservative bound.

\subsection{Blue Model -- Quadratically Lifted Flat Direction}\label{ModBlue}

In the seminal work~\cite{Linde}, hybrid inflation is implemented by a
quadratically lifted
flat direction, through the effective potential
\begin{equation}
\label{VL:blue}
V_{\rm L}(\sigma)=\frac 12 m_\sigma^2 \sigma^2\,.
\end{equation}
Due to the positive curvature of the potential along the flat direction,
the scalar perturbations are predicted to be blue tilted, which is
characterized by a scalar spectral index $n_{\rm s}>1$.
Using~(\ref{Ne}) and the basic potential~(\ref{hybpot}), we can solve for
\begin{equation}
\label{se:blue}
\sigma_{\rm e}=\frac mg 
\exp\left\{\frac{\lambda}{2\pi} \frac{m_{\rm Pl}^2 m_\sigma^2}{m^4} N_{\rm e}\right\}
\,,
\end{equation}
while the amplitude of the power spectrum~(\ref{SPR}) is given by
\begin{equation}
\sqrt{P_{\cal R}}=
\sqrt{\frac 23 \pi} \frac{g m^5}{\lambda^{3/2}m_{\rm Pl}^3 m_\sigma^2}\,.
\end{equation}

The latter two equations can be solved for $m_\sigma$ and $\sigma_{\rm e}$
by assuming that the exponent in~(\ref{se:blue}) is small,
approximating $\sigma_{\rm e}\approx \sigma_{\rm c}$, and justifying this
a posteriori.
We find
\begin{equation}
\label{msblue}
m_\sigma^2 =\frac{g}{\lambda^{3/2}}
\sqrt{\frac{2\pi}{3 P_{\cal R}}}
\frac{m^5}{m_{\rm Pl}^3}\,,
\end{equation}
and
\begin{equation}
\label{seblue}
\sigma_{\rm e}=\frac mg 
\exp\left\{\frac{g}{\sqrt{6\pi\lambda P_{\cal R}}}\frac m{m_{\rm Pl}} N_{\rm e}
\right\}\,,
\end{equation}
such that
\begin{equation}
\label{Deltasblue}
\Delta \sigma\approx \frac{1}{\sqrt{6\pi\lambda P_{\cal R}}}\frac {m^2}{m_{\rm Pl}} N_{\rm e}
\end{equation}
Inserting these into~(\ref{SENum}) and using~(\ref{SPRnum}) yields
\begin{equation}
\label{SEblue}
S_{\rm E}=\frac{42}{g\sqrt{\lambda P_{\cal R}}}\frac{m}{m_{\rm Pl}}N_{\rm e}
=\frac{9.3\times 10^5}{g\sqrt{\lambda}}\frac{m}{m_{\rm Pl}}N_{\rm e} \,.
\end{equation}

We now discuss the self-consistency of the above results. For the
approximation of the potential $V$
by expression~(\ref{V:approx}) to be valid for the bounce solution,
we have to fulfill the relation~(\ref{assume2}) with $w=w_{\rm r}$.
Using~(\ref{wr}) and~(\ref{Deltasblue}) with
$N_{\rm e}=60$, we find the bound
\begin{equation}
\label{scboundblue}
m\ll 2.2\times 10^{-6} \frac g{\sqrt{\lambda}} m_{\rm Pl}\,.
\end{equation}
This condition also ensures the validity of the assumption
$\Delta\sigma\ll \sigma_{\rm c}$, in particular that the exponent
in~(\ref{seblue}) is much smaller than one.

\begin{figure}[htbp]
\begin{center}
\epsfig{file=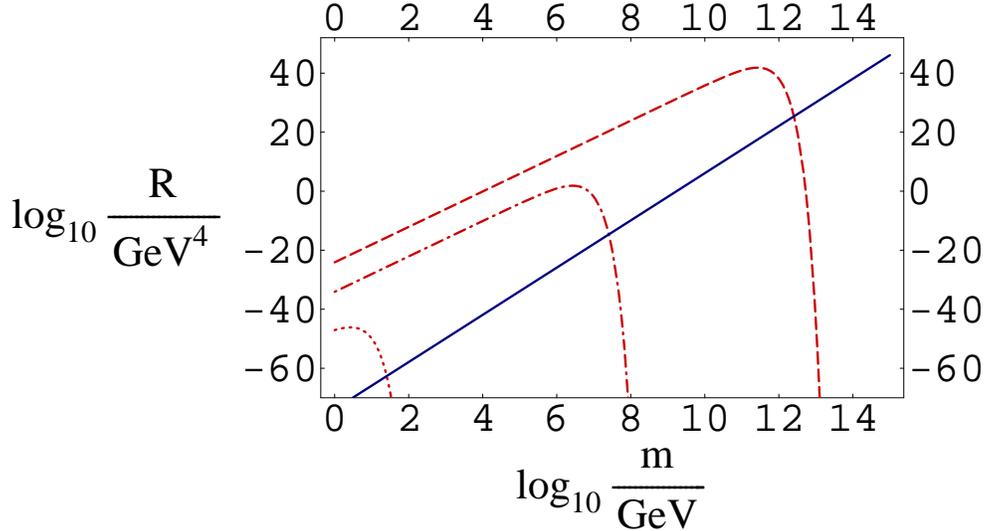, height=3.2in 
       }
\end{center}
\caption{
\label{figureblue}
\small
Hubble expansion {\it vs.} tunneling decay in the blue model. The plot shows
$\log_{10} R$ for $R=H^4$ (solid) and $R=\Gamma/{\cal V}$ for $g=0.5$ (dashed),
$g=5\times 10^{-6}$ (dot-dashed) and $g=5\times 10^{-12}$ (dotted). We have
chosen $\lambda=0.5$.
}
\vskip -0.in
\end{figure}


In order to summarize these results, we present Fig.~\ref{figureblue}.
A reasonable estimate of the tunneling rate is given by
\begin{equation}
\frac{\Gamma}{\cal V}=\frac{S_{\rm E}^2}{4\pi^2}g^4 m^4{\rm e}^{-S_{\rm E}},
\end{equation}
since $g m$ is the smallest dimensionful scale occurring in the approximate
potential~(\ref{V:approx}).
We compare the decay rate with the Hubble rate~(\ref{Hubble}),
since $(\Gamma/{\cal V})/H^4$ is the number of bubbles nucleating in one Hubble
time within a Hubble volume.
Note that for the range of $m$ for the individual graphs
of $\Gamma/{\cal V}$ in Fig.~\ref{figureblue},
the consistency condition~(\ref{scboundblue}) is met.
The wide range of orders of magnitude
covered relativizes the importance of the
prefactor of the exponential in the expression for $\Gamma/{\cal V}$,
in particular the determinants in
\eq~(\ref{tunnelingrate}). Also the precise bound on the tunneling rate
loses importance due to its strong dependence on
$m$ after it has reached is maximum. As a conservative requirement,
we may impose $(\Gamma/{\cal V})<H^4$. A bound which is stronger by a few
orders of magnitude might be in order to accord with observation, but has
no significant impact on the tunneling bound on $m$.


An important implication to be read off from Fig.~(\ref{figureblue}) is that
for the TeV-scale as
a special scale of interest, $g$ has to be smaller than at most $10^{-11}$
in order to avoid a fast end of inflation through tunneling, provided
$\lambda$ is of order one. Besides
by suppressing $g$, we see from \eq~(\ref{SEblue}), that also small
values of $\lambda$ serve to suppress the tunneling rate. Choosing this
option however leads to expectation values for $\phi$ after inflation
and $\sigma$ during inflation, which are much larger than $m$. If one considers
$m$ as a cutoff scale or to be closely related to a cutoff scale of an
effective theory, this
is undesirable.

As a curiosity, we note that we rule out a particular choice of
parameters used as an example in the original work on hybrid
inflation~\cite{Linde},
$m=1.3\times 10^{11} \, {\rm GeV}$, $g^2=\lambda =0.1$. In this case,
$\Gamma/{\cal V}=1.3\times 10^{33}\, {\rm GeV}^4$,
whereas $H^4=1.7\times 10^{14} \,{\rm GeV}^4$, indicating that
$\Gamma/({\cal V} H^4)=7.6\times 10^{18}$ non-inflationary bubbles are
nucleated during one expansion time within a horizon.

\subsection{Red Model}

\label{secRedModel}

Since the WMAP3 data strongly prefers a red-tilted scalar spectral index
$n_{\rm s}$, with the best-fit value given by
$n_{\rm s}\approx 0.95$~\cite{WMAP3}, we also study models with
a negative curvature along the flat direction.
A simple possible realization of these is given by
\begin{equation}
\label{VL:red}
V_{\rm L}=A^3 \sigma - \frac 12 m_\sigma^2 \sigma^2\,.
\end{equation}
During inflation, the inflaton takes values in between $\sigma_{\rm c}$
and the maximum of $V_{\rm L}$, which is located at
$\sigma=A^3/m_\sigma^2$. This translates into the requirement
\begin{equation}
\frac mg < \frac{A^3}{m_\sigma^2}\,.
\end{equation}

Note that this model has an additional parameter when compared with the
quadratically lifted model, which is fixed by imposing the value of
the spectral
index of the scalar perturbations $n_{\rm s}=0.95$
as an additional constraint. It is calculated through
the slow-roll parameter $\eta$ as
\begin{equation}
n_{\rm s}=1+2\eta\,,
\end{equation}
where
\begin{equation}
\eta=\frac{m_{\rm Pl}^2}{8\pi} \frac{\partial^2 V/\partial \sigma^2}{V}\,.
\end{equation}

Imposing the spectral index constraint together with equations~(\ref{Ne})
and~(\ref{SPR}), we find the relations
\begin{eqnarray}
\label{msred}
m_\sigma^2&=&-\eta \frac{2\pi}{\lambda}\frac{m^4}{m_{\rm Pl}^2}\,,\\
A^3&=&- \frac{2\pi \eta}{g \lambda}\frac{m^5}{m_{\rm Pl}^2}
+\sqrt{\frac{2\pi}{3 P_{\cal R}}}\lambda^{-3/2}
\frac{m^6}{m_{\rm Pl}^3}{\rm e}^{-\eta N_{\rm e}}\,,\\
\sigma_{\rm e}&=&\frac mg +  
\frac{1-\exp\{-\eta N_{\rm e}\}}{\eta \sqrt{6\pi \lambda P_{\cal R}}}
\frac{m^2}{m_{\rm Pl}}\,.
\end{eqnarray}

With the numerical result for the Euclidean action~(\ref{SENum})
and the power spectrum normalization~(\ref{SPRnum}), this gives
\begin{equation}
S_{\rm E}=\frac{42}{\eta g \sqrt{\lambda P_{\cal R}}}
\left(1-\exp\{-\eta N_{\rm e}\}\right)\frac{m}{m_{\rm Pl}}
=\frac{9.3\times 10^5}{\eta g \sqrt{\lambda}}
\left(1-\exp\{-\eta N_{\rm e}\}\right)\frac{m}{m_{\rm Pl}}
\,,
\end{equation}
and,  when additionally imposing $N_{\rm e}=60$, $\eta=-0.025$,
\begin{equation}
\label{SEred}
S_{\rm E}=\frac{1.3\times 10^8\, m}{g\sqrt{\lambda}m_{\rm Pl}}\,.
\end{equation}
The consistency condition~(\ref{assume2}) with $w=w_{\rm r}$
for our approximation is fulfilled when
\begin{equation}
m \ll 9.7\times 10^{-7} \frac g{\sqrt{\lambda}} m_{\rm Pl}\,.
\end{equation}

Again, we have found that tunneling is preferred for large couplings
$\lambda$ and $g$ and small values for the mass parameter $m$, where the
small ratio to the Planck scale is imposed by the small amplitude
of density perturbations.
Comparison of the Euclidean actions for the blue model~(\ref{SEblue})
and the red~(\ref{SEred}) with $N_{\rm e}=60$ shows that both differ
only by a proportionality factor which is irrelevant with respect to
the level of our approximation. The figure for the blue model
is therefore almost indistinguishable for the eye when compared
to Fig.~\ref{figureblue} for the red model, which is why it is omitted here.

We note that for the above models, one should bear in mind that
in order to obtain the effective potentials~(\ref{VL:blue}) and~(\ref{VL:red})
in the parametric range which allows for tunneling decay,
one has to require a more than substantial amount of
tuning. For hybrid inflation at the Electroweak scale, this is discussed
in~\cite{Lyth,CoLyRaTr}.
The one-loop correction to the tree-level hybrid
potential~(\ref{hybpot}) is given by
\begin{equation}
\label{V1loop}
V_{1-{\rm loop}}=\frac 1{64\pi^2} \left(m^2 -g^2 \sigma^2\right)^2
\log \frac{g^2 \sigma^2 -m^2}{\Lambda^2}\,,
\end{equation}
where $\Lambda$ is an ultraviolet cutoff. Suppose now, we choose the
renormalizable counterterms, which are the terms up to fourth order in
$\sigma$,
in such a way that for $\sigma=\sigma_{\rm e}$,
we have the desired values for $m_\sigma$ and $A$, while the cubic and quartic
self-couplings cancel to zero. When now expanding the potential around
$\sigma_{\rm e}$ down to values of
$\sigma \stackrel{>}{{}_\sim} \sigma_{\rm c}$, the
nonrenormalizable fifth order term, which we
did not eliminate, induces an additional mass for the field $\sigma$ of
order $g m$. This is to be compared with $m_\sigma$ as in~(\ref{msblue})
or~(\ref{msred}). In order for the quantum correction to be subdominant,
one therefore has to require $g\ll m^3/m_{ \rm Pl}^3$ (blue) or
$g\ll m^2/m_{ \rm Pl}^2$ (red), respectively.
Comparing with Fig.~\ref{figureblue}, it is easy to see that tunneling does not
play any role within hybrid models which do not require the fine-tuning
of nonrenormalizable operators.
This reasoning already strongly indicates that the supersymmetric
scenarios with radiatively lifted flat directions
do not suffer from tunneling either, as we shall see explicitly
in the next section, number~\ref{SUSYhybrid}.

The above study shows that inflation exit via tunneling is mostly
relevant for inflation models with an energy scale below the GUT scale
and especially for the intriguing case of models 
where inflation is connected to electroweak physics and therefore
within experimental reach. In fact, a hybrid model has been suggested
where the role of the waterfall field $\phi$ is played by the Standard Model
Higgs field, such that the field content only needs to be extended by the
inflaton singlet $\sigma$~\cite{GaGrKuSh,KraussTrodden}.
Since these models apparently also bear the
potential for successful baryogenesis, the enormous fine
tuning of the potential may be considered worth the price for a minimal
field content. However, our analysis shows that for the desired
parameters $m/ \sqrt{\lambda} = 246 \, {\rm GeV}$, and the
couplings $g,\;\lambda={\cal O}(1)$ in order
to allow for strong pre- or reheating,
a rapid decay via tunneling is inevitable, such that the electroweak hybrid
models are not viable, even if fine tuned.

Another motivation for contemplating low scales of inflation originates
from models
explaining the hierarchy between the Planck scale and a fundamental unified
scale by the presence of large extra dimensions. Inflation then has
to take place at energies below the unified scale.
Kaloper and Linde~\cite{KaloperLinde} point out that hybrid inflation is
a possible way to keep the vacuum expectation value of the scalar fields in the
effective four-dimensional potential for inflation below the fundamental scale.
While they agree with Lyth on the view that when inflation occurs at the
Electroweak scale,
one faces a severe fine tuning problem, they conclude that models with
$M\sim 10^{11} \,{\rm GeV}$, $g^2=\lambda=0.1$ fit perfectly
in the hybrid scenario, somewhat in contradiction with our estimate
$g\ll m^3/m_{ \rm Pl}^3$. Note also our comment on the tunneling rate
for this parametric range at the end of section~\ref{ModBlue}.

\subsection{SUSY Hybrid Inflation}
\label{SUSYhybrid}

From the general arguments on the irrelevance of the tunneling rate
within our approximation for models without fine-tuning of nonrenormalizable
operators, it is already clear that tunneling does not play a role
in SUSY-hybrid inflation~\cite{Copeland:1994vg,Dvali:1994ms}. These models
are however of special interest since they rely on rather minimal assumptions
and in their simplest version depend only on a single parameter
$\kappa$~\cite{Senoguz:2003zw},
which can be determined~\cite{Battye:2006pk} from the latest observational
data~\cite{WMAP3}~\footnote{Before the WMAP3 data became available, only an
upper bound on $\kappa$ could be given.}.
They furthermore bear the potential of
a successful embedding of the Minimal Supersymmetric Standard Model in
an inflationary scenario, possibly linked to a Grand Unified
Theory~\cite{Dvali:1994ms,Kyae:2005vg,Garbrecht:2005rr,Garbrecht:2006ft}.
Due to the importance of these models,
we derive here an expression for the Euclidean bounce action $S_{\rm E}$,
although it will be large and prohibit tunneling.

$F$-term SUSY hybrid inflation is implemented by the superpotential
\begin{equation}
W=\kappa S \left(G \overline G -M^2\right)\,,
\end{equation}
which leads to the tree-level scalar potential
\begin{equation}
\label{VSUSY}
V=\kappa^2|G\overline G -M^2|^2+\kappa^2|S G|^2+\kappa^2 |S \overline G|^2
\,.
\end{equation}
The involved fields are complex, where $S$ is a singlet, $G$ a gauged
multiplet of
dimension ${\cal N}$ and $\overline G$ its conjugate. Vanishing of the
$D$-terms relates the vacuum expectation values
$\langle G\rangle =\langle \overline G\rangle^*$.

We choose the phase of $S$ to be zero and identify
\begin{equation}
\sigma=\sqrt 2 {\rm Re}[S]\,.
\end{equation}
Likewise, by a unitary gauge choice, such that
$\langle {\rm Re}[G^i] \rangle =\langle |G| \rangle$, we can set
\begin{equation}
\phi=\sqrt 2 {\rm Re}[G^i]\,.
\end{equation}
In terms of these fields, the potential~(\ref{VSUSY}) reads
\begin{equation}
V=\kappa^2 M^4 - \kappa^2 M^2 \phi^2 +\frac 14 \kappa^2 \phi^4
+\frac 12 \kappa^2 \sigma^2 \phi^2\,.
\end{equation}
This is a special case of the more general potential~(\ref{hybpot})
with the replacements
\begin{eqnarray}
m^2 &=& 2\kappa^2 M^2\,,\nonumber\\
g &=& \kappa\,,\nonumber\\
\lambda &=& \kappa^2\,.
\end{eqnarray}
The lifting of the flat direction is then induced
by the Coleman-Weinberg potential~\cite{Dvali:1994ms,Garbrecht:2006ft}
\begin{eqnarray}
V_{\rm L}&=&\frac{\cal N}{32 \pi^2}\kappa^4
\left\{
\left(\frac{\sigma^2}{2}+M^2\right)^2
\log\left( \kappa^2 \frac{\frac 12 \sigma^2 +M^2}{\Lambda^2}\right)
+\left(\frac{\sigma^2}{2}-M^2\right)^2
\log\left( \kappa^2 \frac{\frac 12 \sigma^2 -M^2}{\Lambda^2}\right)
\right.
\nonumber\\
&&
\left.
-\frac 12 \sigma^4 \log\left(\kappa^2 \frac 12 \frac{\sigma^2}{\Lambda^2}\right)
\right\}
\,.
\end{eqnarray}

We consider again the situation where $\sigma$ is close to the critical
point, such that we can approximate
\begin{equation}
\frac{\partial V_{\rm L}}{\partial \sigma}
\approx
\sqrt 2 \log 2 \frac{\cal N}{8\pi^2}\kappa^4 M^3 + O(M^2 \Delta \sigma)\,,
\end{equation}
where the critical point is at $\sigma=\sigma_{\rm c}=\sqrt 2 M$ and
$\Delta \sigma =\sigma -\sigma_c$.
Within this approximation, the number of e-folds~(\ref{Ne}) is
\begin{equation}
N_{\rm e}= \frac{64 \pi^3}{\sqrt 2 \log 2}
\frac{M \Delta \sigma}{\kappa^2 {\cal N} m_{\rm Pl}^2}\,.
\end{equation}
Imposing the normalization of the power spectrum~(\ref{SPR}),
we get a relation between $\kappa$ and $M$,
\begin{equation}
M=\kappa^{1/3}\left(\frac{\sqrt{3\,P_{\cal R}} \, {\cal N} \log 2}{\pi^{5/2}}\right)^{1/3}
\frac{m_{\rm Pl}}4\,,
\end{equation}
such that we can derive
\begin{eqnarray}
\Delta \sigma = (\log 2)^{1/3} 2^{-3/2} 3^{-1/3} \pi^{-4/3}
\kappa^{4/3} \left(\sqrt{P_{\cal R}}\right)^{-2/3} M {\cal N}^{1/3} N_{\rm e}\,.
\end{eqnarray}
Using~(\ref{SENum}), we find for the Euclidean tunneling action
\begin{equation}
S_{\rm E}=6.1 \left(\kappa \sqrt{P_{\cal R}}\right)^{-2/3} {\cal N}^{1/3} N_{\rm e}
=4800 \kappa^{-2/3}{\cal N}^{1/3} N_{\rm e}\,.
\end{equation}
Tunneling therefore does not occur within $F$-term SUSY-hybrid inflation.

We have also performed a corresponding study for the $D$-term
model~\cite{Halyo:1996pp,Binetruy:1996xj},
which is more involved due to the additional parametric dependence on the
gauge coupling constant. However, as one can already anticipate from
the general arguments about radiative corrections and tunneling given at
the end of section~\ref{secRedModel}, we find that tunneling is also very
suppressed in these scenarios. We therefore omit a detailed presentation
of the derivation of this negative result.

\section{Conclusions}

Imposing the normalization of the scalar perturbation
spectrum~(\ref{SPR}), it is possible to estimate for generic
models of hybrid inflation the range of parameters where tunneling
dominates over the slow-roll evolution of the inflaton fields.
In order to calculate the Euclidean action $S_{\rm E}$, we have assumed
that the bounce solution follows a straight trajectory in the field space
spanned by the inflaton and the waterfall field. We have
numerically obtained the action for a particular set of parameters and
then used its scaling properties
in order to calculate the tunneling rates
in parametric regions of small couplings and small field values,
which are difficult to access numerically. This result is expressed in
\eq~(\ref{SENum}), which we have used to derive constraints
on hybrid inflation, arising from the requirement that tunneling should
be suppressed. The consistency of our approach is
verified by a comparison with the numerically determined results for the
bounce action along the curved extremal path in two-dimensional
field space.

Our results are best summarized by the formulas~(\ref{SEblue}), (\ref{SEred})
and by Fig.~\ref{figureblue}. Tunneling may play a role for models with
a mass below $10^{12}\,{\rm GeV}$, but can effectively be suppressed by
small values of the inflaton-waterfall coupling $g$ and the
waterfall self coupling $\lambda$, which in turn imply large expectation
values of the inflaton field during inflation or the waterfall field after
its end.

Provided one does not allow for the fine-tuning of nonrenormalizable operators,
tunneling never constitutes a problem. In particular, one cannot
derive any tunneling bounds on the parameters of $F$- or $D$-term SUSY
models.
In contrast, models of electroweak hybrid inflation, which need coupling
constants of order one for a sufficient reheating of the Universe but require
fine-tuning, are completely ruled out since the inflaton would rapidly decay
through bubble nucleation.
Leaving alone the issue of
stability of the inflaton potential with respect
to radiative corrections, tunneling decay prohibits
the realization of hybrid inflation when the vacuum
energy and all field expectation values are required to be at scales
below $10^{12}\,{\rm GeV}$. 

\section*{Acknowledgements}

T.K. would like to thank Tommy Ohlsson and Malcolm Fairbairn for useful
discussions.  T.K. is supported by the Swedish Research Council
(Vetenskapsr{\aa}det), Contract No.~621-2001-1611.

\end{document}